# PM2.5 and all-cause mortality

S. Stanley Young, CGStat
Warren Kindzierski, University of Alberta

## Abstract

The US EPA and the WHO claim that PM2.5 is causal of all-cause deaths. Both support and fund research on air quality and health effects. WHO funded a massive systematic review and meta-analyses of air quality and health-effect papers. 1,632 literature papers were reviewed and 196 were selected for meta-analyses. The standard air components, particulate matter, PM10 and PM2.5, nitrogen dioxide, NO2, and ozone, were selected as causes and all-cause and cause-specific mortalities were selected as outcomes. A claim was made for PM2.5 and all-cause deaths, risk ratio of 1.0065, with confidence limits of 1.0044 to 1.0086. There is a need to evaluate the reliability of this causal claim. Based on a p-value plot and discussion of several forms of bias, we conclude that the association is not causal.

## Introduction

The US EPA and the WHO claim that PM2.5 is causal of all-cause deaths. Both support and fund research on air quality and health effects. WHO funded a massive systematic review and meta-analysis of air quality and health effects papers, Orellano (2020ab), hereafter Orellano. In Orellano, 1,632 literature papers were identified and reviewed and 196 were selected for meta-analyses. The standard air components, particulate matter (PM10 and PM2.5), nitrogen dioxide (NO2), and ozone (O3), were selected as "causes" and all-cause and cause-specific mortalities were selected as outcomes.

### Data

Orellano gives multiple meta-analyses that suggest many causal claims. For example, they claim that PM2.5 is causal of all-cause deaths. They give 29 log risk ratios and log Standard Errors taken from 27 papers, Orellano (2020 ab), for all-cause mortality and PM2.5. See Table 1; the data was taken from Orellano Supplemental material A.5.

### Methods: P-value plot

Briefly, p-values were computed, ranked from smallest to largest and plotted against the integers to give a p-value plot. See Young and Kindzierski (2018) for p-value plot formation details.

### Methods: Bias and small p-values

If there is a constant bias, B, regardless of the sample size n, then there is a p-value problem as n gets large. This is explained further. The measured value is M and it is equal to the true value T of a parameter plus any bias, B; M = T + B. So, if we test M with larger and larger samples, and if the real value of T is zero, then T will converge to zero and we will be left with B. As the Standard Error, SE, gets small, B/SE, gets large and we can get an extremely small p-value.

**Results**

Finally, as the authors provide risk ratios with confidence limits, it is easy enough to produce a p-value plot. Using the data in Table 1, we present our p-value plot for all-cause mortality and PM2.5 in Figure 1.

In Figure 1 we see the rather common bilinear p-value plot. There are 13 p-values below 0.05 and 16 p-value above 0.05.

**Discussion**

Orellano claims a risk ratio of 1.0065, with 95% confidence limits of 1.0044 to 1.0086. Keep in mind that a risk ratio of 1.000 is considered no effect. So far as we know, 1.0065 is the smallest risk ratio claim for PM2.5 in the literature.

A causal claim can be disputed on several grounds. First, any small bias in the selection of base papers to use in the meta-analysis might tilt the risk ratio. The papers that were candidates for selection as base papers for the meta-analysis might themselves be a biased set – publication bias; negative papers rarely make it into print. The base papers might not have corrected for multiple testing or multiple modeling; a base paper with a small random p-value might pass a biased risk ratio to the meta-analysis, Young and Kindzierski (2019). With a risk ratio this small, very slight biases might lead to the Orellano result. Given our counting of questions/models in other environmental epidemiology papers, Young and Kindzierski (2019), it is reasonable to think the numbers of questions/models in base papers here are no better than those that have been examined where the median number of questions/models was on the order of 10,000.

Exceedingly small p-values require explanation. The smallest p-value from Table 1 is $3.27 \times 10^{-19}$, a value so small as to imply certainty. A p-value this small can come about by bias and a small Standard Error. What are some possible sources of bias? High or low temperature can kill the old and the weak. Living in a controlled temperature environment can protect individuals from extreme temperature. There was a heat wave in northern Europe in the year 2003. There were an estimated 70,000 excess deaths as many people did not have air conditioning. Spikes in temperature can kill. High temperatures often occur with high pressure and little wind, and these are conditions that allow poor air quality to develop. PM2.5 can be high while temperature is high.

Young et al. (2017) and You et al. (2018) examined a large California PM2.5/mortality data set, ~37k exposure days, ~2M e-death certificates. The analysis data set for these studies is public. No evidence of PM2.5 associated mortality was found. These two papers were not included in Orellano, which suggests possible selection bias in the selection of their base papers.

Multiple factors – multiple testing and multiple modeling bias, publication bias, bias combined with a small standard error, large negative studies, etc., argue against the Orellano claim being causal. There is no convincing evidence of an effect of PM2.5 on all-cause mortality in their study.

# Tables and Figures

Table 1. Natural log of the environmental effect (LnEE) and its Standard Error (SELnEE) for 29 selected papers that looked at PM2.5 and all-cause mortality. The data is taken from A.5 of Orellano (2020), supplemental material.

| RowID | ID[1] | Author | Year | LnEE | SELnEE | Z | p-value | Rank |
|---|---|---|---|---|---|---|---|---|
| 1 | 337 | Dai | 2014 | 0.011731 | 0.001309 | 8.959154 | 3.27E-19 | 1 |
| 2 | 273 | Chen | 2011 | 0.004589 | 0.000558 | 8.219683 | 2.04E-16 | 2 |
| 3 | 758 | Lee | 2016 | 0.01548 | 0.001905 | 8.123914 | 4.51E-16 | 3 |
| 4 | 772 | Li | 2017 | 0.001699 | 0.000306 | 5.559715 | 2.7E-08 | 4 |
| 5 | 633 | Janssen | 2013 | 0.007968 | 0.002021 | 3.943442 | 8.03E-05 | 5 |
| 6 | 1774 | Li | 2018 | 0.002497 | 0.000661 | 3.776385 | 0.000159 | 6 |
| 7 | 1409 | Tsai | 2014 | 0.039268 | 0.010868 | 3.613243 | 0.000302 | 7 |
| 8 | 851 | Madsen | 2012 | 0.027615 | 0.00788 | 3.504571 | 0.000457 | 8 |
| 9 | 1733 | Wu | 2018 | 0.005485 | 0.001571 | 3.492337 | 0.000479 | 9 |
| 10 | 1714 | Reyna | 2012 | 0.008243 | 0.002527 | 3.261944 | 0.001107 | 10 |
| 11 | 489 | Garret | 2011 | 0.006678 | 0.002477 | 2.695497 | 0.007028 | 11 |
| 12 | 601 | Hong | 2017 | 0.112056 | 0.049633 | 2.257704 | 0.023964 | 12 |
| 13 | 245 | Castillejos | 2000 | 0.014692 | 0.007387 | 1.988796 | 0.046724 | 13 |
| 14 | 211 | Burnett | 2004 | 0.005993 | 0.003191 | 1.877872 | 0.060399 | 14 |
| 15 | 1691 | Dockery | 1992 | 0.017059 | 0.009617 | 1.773869 | 0.076085 | 15 |
| 16 | 84 | Atkinson | 2016 | -0.01419 | 0.008234 | -1.72311 | 0.084868 | 16 |
| 17 | 1691 | Dockery | 1992 | 0.022793 | 0.018614 | 1.224494 | 0.220766 | 17 |
| 18 | 975 | Neuberger | 2007 | 0.004988 | 0.005052 | 0.987326 | 0.323483 | 18 |
| 19 | 1709 | Simpson | 2000 | 0.007997 | 0.00815 | 0.981175 | 0.326506 | 19 |
| 20 | 974 | Neuberger | 2013 | 0.003992 | 0.004553 | 0.876757 | 0.380618 | 20 |
| 21 | 1695 | Peters | 2009 | 0.004888 | 0.006454 | 0.757427 | 0.448794 | 21 |
| 22 | 1411 | Tsai | 2014 | 0.007692 | 0.012302 | 0.625253 | 0.531805 | 22 |
| 23 | 748 | Lanzinger | 2016 | -0.00404 | 0.006971 | -0.57993 | 0.561965 | 23 |
| 24 | 59 | Anderson | 2001 | 0.00338 | 0.005955 | 0.567519 | 0.570362 | 24 |
| 25 | 827 | Lopez-Villarrubia | 2010 | -0.00914 | 0.016166 | -0.56548 | 0.571746 | 25 |
| 26 | 827 | Lopez-Villarrubia | 2010 | -0.00682 | 0.016925 | -0.40314 | 0.686842 | 26 |
| 27 | 186 | Branis | 2010 | -0.002 | 0.006098 | -0.3283 | 0.742686 | 27 |
| 28 | 120 | Basagana | 2015 | 0.001112 | 0.005225 | 0.212729 | 0.831538 | 28 |
| 29 | 722 | Kollanus | 2016 | 0.001998 | 0.041073 | 0.048645 | 0.961202 | 29 |

[1]Orellano accession number.

Figure 1. P-value plot of all-cause mortality and PM2.5 The data is from the A.5 supplement to Orellano (2020b).

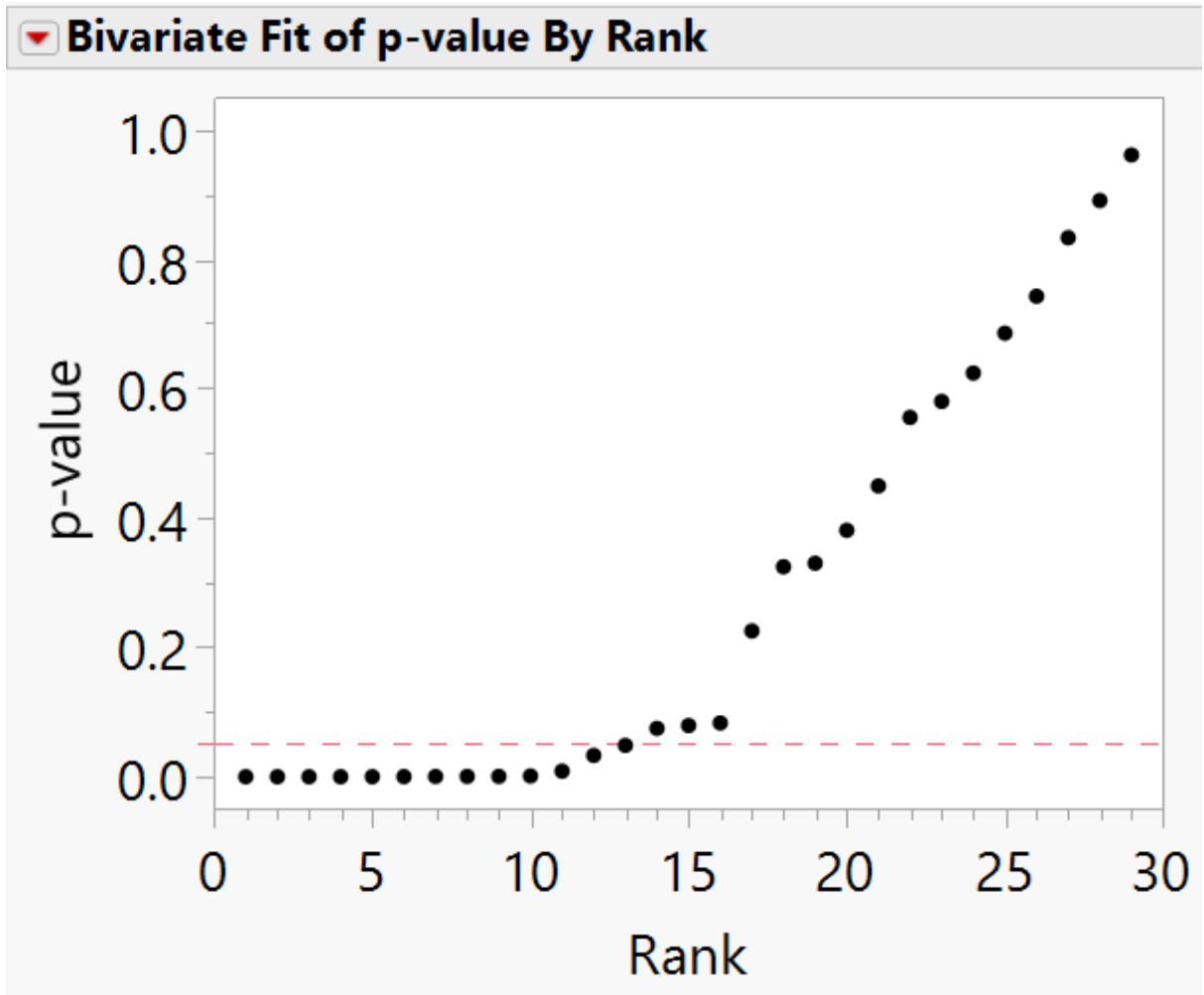